# The evolution of the clustering of QSOs

F. La Franca[1], P. Andreani[2] and S. Cristiani[2]

[1] Dipartimento di Fisica, Università degli studi "Roma Tre", Via della Vasca Navale 84, Roma, I-00146
[2] Dipartimento di Astronomia, Università degli studi di Padova, Vicolo dell'Osservatorio 5, Padova, I-35122

**Abstract.** The evolution of QSO clustering is investigated with a new sample of 388 QSOs with $0.3 < z \leq 2.2$, $B \leq 20.5$ and $M_B < -23$. Evidence is found for an increase of the clustering amplitude with increasing redshift. These measurements allow to further distinguish among the various physical scenarios proposed to interpret the QSO phenomenon. A single population model is inconsistent with the observations. The general properties of the QSO population would arise naturally if quasars are short-lived events connected to a characteristic halo mass $\sim 5 \cdot 10^{12}$ $M_\odot$.

## 1 Introduction

The first detections of the quasar clustering date back more than one decade (Shaver 1984). Up to now, however, more detailed studies of the clustering dependence on physical parameters like absolute magnitude and redshift was hampered by the small number of quasars in statistically well-defined samples. In recent times complete samples totaling about 2000 QSOs have been used resulting in a $4 - 5\sigma$ detection of the clustering on scales of the order of $6h^{-1}$ comoving Mpc (Andreani & Cristiani 1992, Mo & Fang 1993, Croom & Shanks 1996). The evolution of this clustering is not clear. An amplitude constant in comoving coordinates or marginally decreasing with increasing redshift has been suggested, an amplitude which appears to be consistent or slightly larger than what is observed for present-day galaxies and definitely less than the clustering of clusters.

## 2 Methods and results

In an attempt to improve the situation, while waiting for the 2dF QSO redshift survey, we have carried out a survey in the South Galactic Pole (SGP) over a *connected* area of 25 square degrees down to $B_j = 20.5$ (La Franca et al. 1998). Stacked UKSTU plates were used to select UVx candidates and the multi-fiber spectrograph MEFOS at ESO to take spectra of them. The final sample is made up of 388 QSOs with $0.3 < z < 2.2$. The data set was divided into several luminosity, redshift and spatial sub-samples in order to study the autocorrelation function $\xi(r)$ and the integral autocorrelation function $\bar{\xi}(r)$ as a function of the comoving distance, assuming a fixed value of $\gamma = 1.8$. The two point correlation function (TPCF) analysis gives an amplitude $r_o = (6.2 \pm 1.6)\ h^{-1}$ Mpc at an average redshift 1.34. While $\bar{\xi}(25) = 0.21 \pm 0.16$ is found, in agreement with estimate of Croom and Shanks (1996) of $\bar{\xi}(25) = 0.16 \pm 0.08$. However, when the evolution of the clustering with redshift is analyzed, evidence is found for an *increase* of the clustering with increasing redshift (La Franca, Andreani & Cristiani 1998). The sample was split into the two redshift ranges $0.3 < z \leq 1.4$, and $1.4 < z \leq 2.2$ (Fig. 1). These were fitted by $\gamma = 1.8$ power laws with $r_0$ as a free parameter. At low redshift ($z = 0.97$), $r_0 = 4.2\ h^{-1}$ Mpc was found, corresponding to $\bar{\xi}(15) = 0.26 \pm 0.27$; while at high redshift ($z = 1.82$), $r_0 = 9.1\ h^{-1}$ Mpc, which corresponds to $\bar{\xi}(15) = 1.03 \pm 0.36$. The effect is small, a $2\sigma$ significant discrepancy, but it is interestingly corroborated by other results (at lower and higher redshift) in the literature.

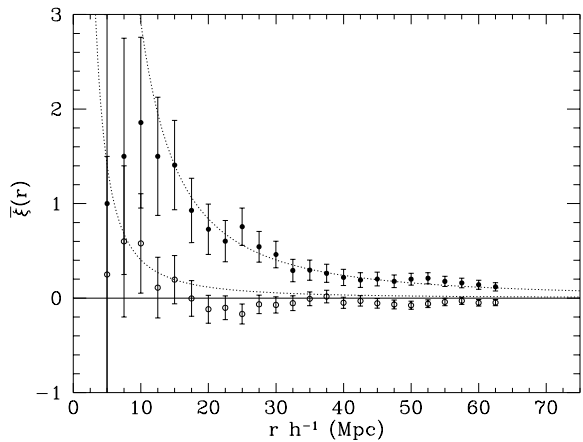

**Fig. 1.** The integral correlation function $\bar{\xi}(r)$ (defined as $\bar{\xi}(r) = \frac{3}{r^3}\int_0^r x^2 \xi(x)dx$) for the quasars in the SGP sample in two redshift ranges $0.3 < z \leq 1.4$, and $1.4 < z \leq 2.2$.

At low redshift Boyle and Mo (1993) measured the clustering of low-$z$ QSOs in the EMSS, while Georgantopoulos and Shanks (1994) used the IRAS point source catalog to measure the clustering of Seyferts. Altogether a low value of the TPCF at 15 Mpc and $z = 0.05$ is obtained, $\bar{\xi} = 0.24 \pm 0.25$. Besides, the data of the Palomar Transit Grism Survey (Kundić 1997, Stephens et al. 1997) allow measuring the amplitude of the TPCF at redshifts higher than 2.7 and the result, $r_o = (18 \pm 8)h^{-1}$ Mpc, suggests that the trend of increasing clustering persists. It may be argued that these surveys tend to select objects with different luminosities and the comparison with the SGP data could not be entirely significant, but an analysis on restricted absolute magnitude slices of the SGP sample shows no correlation of the clustering with the QSO absolute luminosity. If we describe the evolving correlation function in a standard way: $\xi(r,z) = (r/r_0)^{-\gamma}(1+z)^{-(3-\gamma+\epsilon)}$, where $\epsilon$ is an arbitrary (and not very physical,

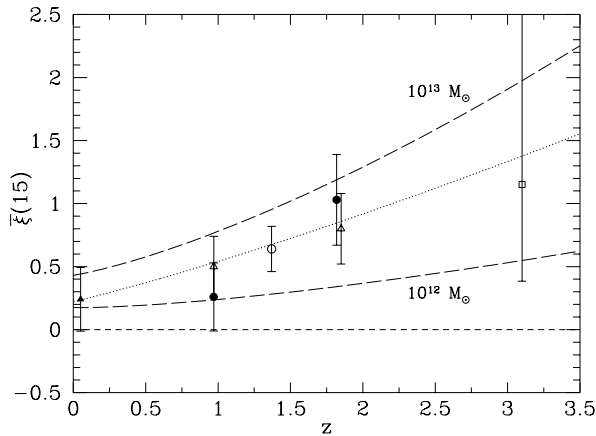

**Fig. 2.** The amplitude of the $\bar{\xi}(15\ h^{-1}$ Mpc) as a function of z. Filled circles: the low- and high-$z$ SGP subsamples (filled circles); open circle: the SGP sample plus the Boyle et al. (1990), La Franca, Cristiani and Barbieri (1992), and Zitelli et al. (1992) samples; open triangles: same as open circle but divided in two redshift slices; filled triangle: low-$z$ AGNs from Boyle and Mo (1993) and Georgantopoulos and Shanks (1994); open square: the high-$z$ sample from Kundić 1997. The dotted line is the $\epsilon = -2.5$ clustering evolution fitted to the open triangles and the filled triangle data. The dashed lines are the $10^{12}$ and $10^{13}\ M_\odot\ h^{-1}$ minimum halo masses clustering evolution according to the transient model of Matarrese et al. (1997).

see Matarrese et al. 1997) fitting parameter, we obtain $\epsilon = -2.5 \pm 1.0$ (Fig. 2).

In spite of the statistical uncertainties the measured QSO clustering is able to put interesting constraints on the allowed evolution, being inconsistent with values $\epsilon > 0.0$, such as $\epsilon \simeq 0.8$ observed for faint galaxies at lower redshifts (Le Fèvre et al. 1996, Carlberg et al. 1997, Villumsen et al. 1997). Great care should be exercised however when carrying out this comparison. Are the faint lower-redshift galaxies representative of the same population of galaxies for which recent observations by Steidel et al (1998) show substantial clustering at $z \simeq 3.1$? Are the Lyman-break galaxies progenitors of massive galaxies at the present epoch or precursors of present day cluster galaxies (Governato et al. 1998)?

We already know from energetic arguments that QSOs cannot shine continuously from high redshifts to the present epoch (Cavaliere and Padovani 1989). However the existing models still do not exclude that a single population exists, which after having been formed at a certain epoch, has undergone a recurrent activity with a sequence of active and quiescent periods. But - following Matarrese et al. (1997) and Moscardini et al. (1998) - this scenario would correspond to an object conserving model in which a decrease of the clustering amplitude with redshift is expected. *Thus we can come to the conclusion that the observed increase of the clustering amplitude with redshift is able to rule out a single* *population model for QSOs.*

If we go back to the model in which quasars are associated with interactions, then we may think in terms of clustering of transient objects, which is definitely different from the case of galaxies which, depending on the physical scenario, can be assimilated to the merging model or the object-conserving paradigm of long-lived objects. In this way the observed clustering is the result of the convolution of the true clustering of the mass with the bias and redshift distribution of the objects. If we think of QSOs as objects sparsely sampling halos with $M > M_{\min}$ we may ask what are the typical masses which allow reproducing the observed clustering. In this perspective an increase of the QSO clustering is expected because they are sampling rarer and rarer overdensities with increasing redshift. As we can see from Fig. 2 an $M_{\min} = 10^{12} - 10^{13}\ M_\odot$ would provide the desired amount of clustering and evolution. Similar theoretical results have also been obtained by Bagla (1997).